\documentclass[letter,traditabstract,longauth,letterpaper]{aa}  

\usepackage{graphicx}
\usepackage{epsfig}
\usepackage{txfonts}
\usepackage{natbib}
\bibpunct{(}{)}{;}{a}{}{,}
\usepackage{flafter}
\newcommand{\bulletc}{$1$E$0657{-}56$}

\begin{document}
\titlerunning{First detection of the SZ effect at $\lambda < 650
  \, \mu$m} \title{First detection
  of the Sunyaev Zel'dovich effect increment at $\lambda < 650 \,
  \mu$m\thanks{Herschel is an ESA space observatory with science
    instruments provided by European-led Principal Investigator
    consortia and with important participation from NASA.  It is open
    for proposals for observing time from the worldwide astronomical
    community.  Data presented in this paper were analyzed using ``The
    Herschel Interactive Processing Environment (HIPE),'' a joint
    development by the Herschel Science Ground Segment Consortium,
    consisting of ESA, the NASA Herschel Science Center, and the HIFI,
    PACS and SPIRE consortia.}}

%   \subtitle{I. Overviewing the $\kappa$-mechanism}

   \author{
          M.~Zemcov\inst{\ref{inst3},\ref{inst4}}
          \and
          M.~Rex\inst{\ref{inst1}}
          \and
          T.~D.~Rawle\inst{\ref{inst1}}
          \and
          J.~J.~Bock\inst{\ref{inst3},\ref{inst4}}
          \and
          E.~Egami\inst{\ref{inst1}}
          \and
          B.~Altieri\inst{\ref{inst2}}
          \and
          A.~W.~Blain\inst{\ref{inst3}}
          \and
          F.~Boone\inst{\ref{inst5},\ref{inst7}}
          \and
          C.~R.~Bridge\inst{\ref{inst3}}
          \and
          B.~Clement\inst{\ref{inst6}}
          \and
          F.~Combes\inst{\ref{inst7}}
          \and
          C.~D.~Dowell\inst{\ref{inst3},\ref{inst4}}
          \and
          M.~Dessauges-Zavadsky\inst{\ref{inst8}} 
          \and
          D.~Fadda\inst{\ref{inst9}}
          \and
          O.~Ilbert\inst{\ref{inst6}}
          \and
          R.~J.~Ivison\inst{\ref{inst10},\ref{inst11}}
          \and
          M.~Jauzac\inst{\ref{inst6}}
          \and
          J.-P.~Kneib\inst{\ref{inst6}}
          \and
          D.~Lutz\inst{\ref{inst12}}
%          \and
%          L.~Metcalfe\inst{\ref{inst2}}
          \and
%          A.~Omont\inst{\ref{inst13}}
%          \and
          R.~Pell\'{o}\inst{\ref{inst5}}
          \and
          M.~J.~Pereira\inst{\ref{inst1}}
          \and
          P.~G.~P\'{e}rez-Gonz\'{a}lez\inst{\ref{inst14},\ref{inst1}}
          \and
          J.~Richard\inst{\ref{inst15}}
          \and
          G.~H.~Rieke\inst{\ref{inst1}}
          \and
          G.~Rodighiero\inst{\ref{inst16}}
          \and
          D.~Schaerer\inst{\ref{inst8},\ref{inst5}}
          \and
%          I.~Smail\inst{\ref{inst15}}
%          \and
          G.~P.~Smith\inst{\ref{inst17}}
          \and
          I.~Valtchanov\inst{\ref{inst2}}
          \and
          G.~L.~Walth\inst{\ref{inst1}}
          \and
          P.~van~der~Werf\inst{\ref{inst18}}
          \and
          M.~W.~Werner\inst{\ref{inst4}}
          }

   \institute{California Institute of Technology, Pasadena, CA 91125,
         USA; \email{zemcov@caltech.edu}\label{inst3}
         \and
         Jet Propulsion Laboratory, Pasadena, CA 91109, USA\label{inst4}
         \and
         Steward Observatory, University of Arizona,
         933 N. Cherry Ave, Tucson, AZ 85721, USA\label{inst1}
         \and
         Herschel Science Centre, ESAC, ESA, PO Box 78, Villanueva de
         la Ca\~nada, 28691 Madrid, Spain\label{inst2}
         \and
         Laboratoire d'Astrophysique de Toulouse-Tarbes,
         Universit\'{e} de Toulouse, CNRS, 14 Av. Edouard Belin, 31400
         Toulouse, France\label{inst5}
         \and
         Laboratoire d'Astrophysique de Marseille, CNRS -
         Universit\'{e} Aix-Marseille, 38 rue Fr\'{e}d\'{e}ric
         Joliot-Curie, 13388 Marseille Cedex 13, France\label{inst6}
         \and
         Observatoire de Paris, LERMA, 61 Av. de l'Observatoire, 75014
         Paris, France\label{inst7}
         \and
         Geneva Observatory, University of Geneva, 51, Ch. des
         Maillettes, CH-1290 Versoix, Switzerland\label{inst8}
         \and
         NASA Herschel Science Center, California Institute of
         Technology, MS 100-22, Pasadena, CA 91125, USA\label{inst9}
         \and
         UK Astronomy Technology Centre, Science and Technology
         Facilities Council, Royal Observatory, Blackford Hill,
         Edinburgh EH9 3HJ, UK\label{inst10}
         \and
         Institute for Astronomy, University of Edinburgh, Blackford
         Hill, Edinburgh EH9 3HJ, UK\label{inst11}
         \and
         Max-Planck-Institut f\"{u}r extraterrestrische Physik,
         Postfach 1312, 85741 Garching, Germany\label{inst12}
         \and
%         Institut d'Astrophysique de Paris, CNRS and Universit\'{e}
%         Pierre et Marie Curie, 98bis Boulevard Arago, F-75014 Paris,
%         France\label{inst13}
%         \and
         Departamento de Astrof\'{\i}sica, Facultad de
         CC. F\'{\i}sicas, Universidad Complutense de Madrid, E-28040
         Madrid, Spain\label{inst14}
         \and
         Institute for Computational Cosmology, Department of Physics,
         Durham University, South Road, Durham DH1 3LE, UK\label{inst15}
         \and
         Department of Astronomy, University of Padova,
         Vicolo dell'Osservatorio 3, I-35122 Padova, Italy\label{inst16}
         \and
         School of Physics and Astronomy, University of Birmingham,
         Edgbaston, Birmingham, B15 2TT, UK\label{inst17}
         \and
         Sterrewacht Leiden, Leiden University, PO Box 9513, 2300 RA
         Leiden, the Netherlands\label{inst18}
         }

   \date{Submitted April 1, 2010, accepted May 10, 2010}

  \abstract{The Sunyaev--Zel'dovich (SZ) effect is a spectral
    distortion of the cosmic microwave background as observed through
    the hot plasma in galaxy clusters.  This distortion is a decrement
    in the CMB intensity for $\lambda > 1.3 \,$mm, an increment at
    shorter wavelengths, and small again by $\lambda \sim 250 \,
    \mu$m.  As part of the \textit{Herschel} Lensing Survey (HLS) we
    have mapped \bulletc\ (the Bullet cluster) with SPIRE with bands
    centered at $250$, $350$ and $500 \, \mu$m and have detected the
    SZ effect at the two longest wavelengths.  The measured SZ effect
    increment central intensities are $\Delta I_{0} = 0.097 \pm 0.019
    \,$MJy sr$^{-1}$ at $350 \, \mu$m and $\Delta I_{0} = 0.268 \pm
    0.031 \,$MJy sr$^{-1}$ at $500 \, \mu$m, consistent with the SZ
    effect spectrum derived from previous measurements at $2 \,$mm.
    No other diffuse emission is detected.  The presence of the finite
    temperature SZ effect correction is preferred by the SPIRE data at
    a significance of $2.1 \sigma$, opening the possibility that the
    relativistic SZ effect correction can be constrained by SPIRE in a
    sample of clusters.  The results presented here have important
    ramifications for both sub-mm measurements of galaxy clusters and
    blank field surveys with SPIRE.}

   \keywords{cosmic background radiation -- Galaxies: clusters:
     individual: \bulletc}

   \maketitle

%________________________________________________________________

\section{Introduction}

The Sunyaev-Zel'dovich (SZ) effect is a distortion of the spectral
shape of the cosmic microwave background (CMB) due to inverse Compton
scattering in the ubiquitous, hot ($T_{\mathrm{e}} \sim 10^{7} \,$K)
intracluster medium (ICM) of galaxy clusters \citep{Sunyaev1972}.  The
canonical thermal SZ spectrum is a decrement in the brightness of the
CMB as measured through galaxy clusters at mm wavelengths and an
increment at sub-mm wavelengths which passes though a null at $\lambda
\approx 1.3 \,$mm.  To correctly describe the SZ spectral distortion
when relativistic electrons are present or the cluster is moving with
respect to the CMB additional correction terms, usually termed
``relativistic'' (or ``finite temperature'') corrections and
``kinetic'' SZ effect, are required.  Measurement of these corrections
is only possible using observations at multiple wavelengths, and is
expedited by measurement at wavelengths where the expected
modifications to the thermal SZ effect are largest.  In the case of
the finite temperature corrections, the largest changes expected in
the SZ increment are at wavelengths shorter than $1 \, \mu$m.

Decrements in emission are rare astrophysically and can be ascribed to
the SZ effect with little ambiguity; this has lead to measurements of
the SZ effect at $\lambda \gtrsim 2 \,$mm becoming almost routine.  In
comparison, measurements of the SZ effect increment are complicated by
the presence of the dusty, high redshift galaxies which constitute the
sub-mm cosmic background.  Additionally, the individual sources
comprising the sub-mm background are gravitationally lensed by galaxy
clusters, the effect of which is to preferentially correlate increases
in sub-mm emission with clustering.  This correlation makes
unambiguous detection of the SZ increment difficult, though successful
measurements do exist (\citealt{Lamarre1998}, \citealt{Komatsu1999},
\citealt{Zemcov2003}, \citealt{Zemcov2007}, \citealt{Nord2009}).
Moreover, the presence of the sub-mm background may contaminate
measurements of the SZ effect for $\lambda > 1 \,$mm in less massive
galaxy clusters \citep{Aghanim2005}.  A better understanding of the
sub-mm emission associated with galaxy clusters is required.

Heretofore, systematic far infrared (FIR) surveys of many galaxy
clusters to large radii have been technically challenging so a
complete census of sub-mm emission from clusters has been difficult to
obtain.  The advent of SPIRE \citep{Griffin2010} on \textit{Herschel}
\citep{Pilbratt2010} has, for the first time, provided the capability
to make deep maps of clusters to large angles on the sky and to use
colour information to separate the different sources of sub-mm
emission present in galaxy clusters.  In addition to gravitationally
lensed background sources \citep{Rex2010}, emission in clusters above
the confused sub-mm background may also comprise emission from
galaxies in the cluster itself \citep{Rawle2010}, as well as truly
diffuse emission from the SZ effect and possibly even cold dust in the
ICM.  SPIRE's ability to separate sources of emission based both on
spatial and spectral information allows the demographics of the sub-mm
emission to be measured.  

In this paper, we use deep SPIRE maps of the $z=0.3$ Bullet cluster
(\bulletc) taken as part of the \textit{Herschel} Lensing Survey (HLS,
P.I.~Egami) at $250$, $350$, and $500 \, \mu$m with $18$, $25$ and
$36$ arcsec resolution to measure the SZ effect and constrain other
diffuse emission associated with the cluster.

%________________________________________________________________

\section{Analysis and results}
\label{S:analysis}

The HLS is a programme to observe 40 massive clusters in the range
$0.1 < z < 1.0$; during \textit{Herschel}'s science demonstration
phase \bulletc\ was observed to full depth.  The specific observation
strategy, low level data reduction and map making processes are
summarized in \citet{Egami2010}; here we begin with the calibrated
flux and noise maps at $250, 350$ and $500 \, \mu$m (Fig.~2 of
\citealt{Egami2010}).

Measurement of dim, extended emission is complicated by the presence
of background sub-mm galaxies the confusion noise from which, at
SPIRE's resolution, is measured to be $\sim 6 \,$mJy in all three
bands for blank fields \citep{Nguyen2010}.  Indeed, in these maps we
measure an RMS noise of $\{6.0,6.0,7.5\} \,$mJy/beam at $250, 350, 500
\, \mu$m, consistent with the combination of confusion noise and
instrument noise expected from the integration time.  Obviously, care
must be taken to disentangle extended emission from confused sources.
To do this, we employ the approach discussed below, the fundamental
tenet of which is to account for and remove all $250 \, \mu$m emission
from the two longer wavelength channels.

\begin{figure*}[!ht]
\centering
\epsfig{file=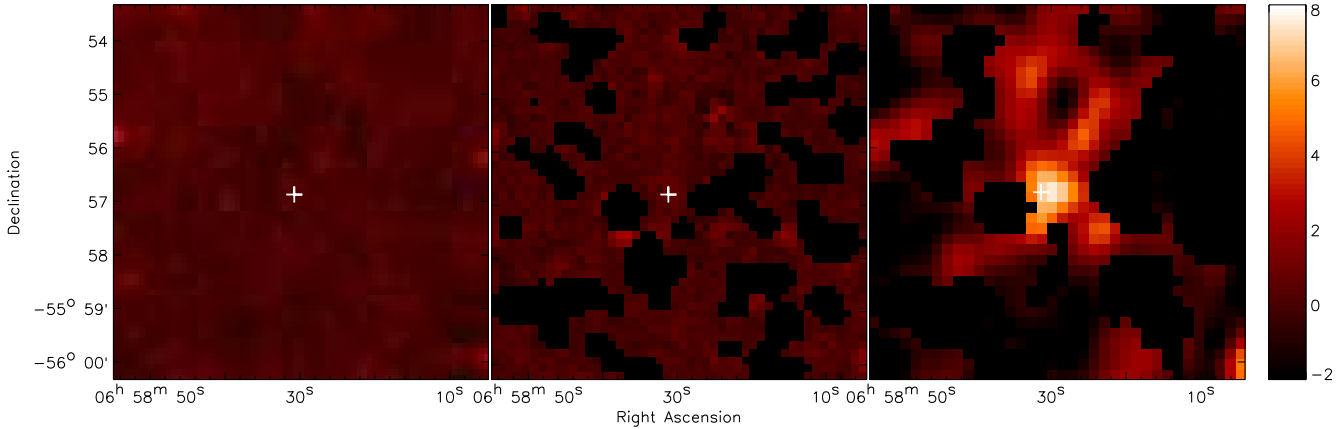,width=0.99\textwidth}
\caption{Source subtracted maps at $250, 350$ and $500 \, \mu$m; these
  figures show the same regions defined by the 7\arcmin$\times
  $7\arcmin green boxes in Fig.~2 of \citet{Egami2010}.  All three
  maps are shown on the colour scale shown at right; units are mJy
  beam$^{-1}$.  The left most panel shows the source-subtracted $250
  \, \mu$m map; the confused sub-mm background is most clearly evident
  in this band.  The center and rightmost panels are respectively the
  $350$ and $500 \, \mu$m source removed and masked maps; their
  construction is discussed in the text.  The cross hairs show the
  best fitting SZ effect centroid from \citet{Halverson2009}; the
  emission visible at $500 \, \mu$m but not in the other maps is
  consistent with the position, shape and flux from the SZ effect
  expected at this wavelength.}
\label{fig:submaps}
\end{figure*}

To generate a $250 \, \mu$m source catalog, sources are detected in
the $250 \, \mu$m map using the StarFinder algorithm
\citep{Diolaiti2000}.  The detection threshold is set to $2.5 \sigma$
and is run iteratively using SPIRE's $250 \, \mu$m point response
function (PRF) to account for blending; as the purpose here is not to
generate a catalog of real sources but rather to remove all potential
sources of emission resolved in the shortest wavelength band, an
aggressive source detection level is desirable.  Because the confusion
noise in these maps is already a factor of $\sim 3$ above the
instrument noise level, lowering the source detection threshold does
not substantially increase the number of identified sources.  The $250
\, \mu$m selected sources are then subtracted from the $250 \, \mu$m
map to produce the source-subtracted map shown in Fig.~\ref{fig:submaps}.
In order to remove $250 \, \mu$m emission that is correlated with
$350$ and $500 \, \mu$m emission we employ a very conservative
approach.  The first step in this process is to use the $250 \, \mu$m
source candidate catalog to determine the possible positions of
counterparts in the other two channels.  The algorithm described below
is performed independently on both the $350$ and $500 \, \mu$m maps;
no direct comparison between the catalogs for the two longer
wavelengths is performed or necessary.  For each $250 \, \mu$m
candidate source position, the relevant PRF is fit to the position in
the longer wavelength channel.  In the fit, the position, flux and
width of the source are allowed to vary.  The resulting list of long
wavelength candidate source counterparts is then compared to the $250
\, \mu$m candidate source catalog.  All those fits where the best fit
flux is different by $> 50$\% at $350 \, \mu$m or $> 90$\% at $500 \,
\mu$m, whose best fit width differs from the nominal PRF by $> 20$\%,
or whose best fitting position is different by a total distance of 2
map pixels or more are deemed ``unsuccessful''.  The catalog
consisting of the successful fits is then subtracted from the long
wavelength map to produce a source subtracted map at that wavelength.

Since the angular resolution of the two longer wavelength channels is
significantly larger than that at $250 \, \mu$m, it is common to have
a large fraction of the fits be unsuccessful.  As this is due to the
effects of confusion rather than an actual lack of a long wavelength
counterpart source, to be conservative we need to account for these
unsubtracted sources; this process is performed in two steps.  First,
we generate a $250 \, \mu$m map which is the sum of the $250 \, \mu$m
source subtracted map (shown in Fig.~\ref{fig:submaps}) and all
those $250 \, \mu$m selected sources for which the longer wavelength
fit was unsuccessful.  This corresponds to the $250 \, \mu$m map which
would arise if all those sources detected and removed at the longer
wavelength were also removed from the $250 \, \mu$m input map.  This
map is then convolved up to the longer wavelength's resolution using
the relevant PRF and rebinned to the longer wavelength map's
resolution.  This map - which is equivalent to the long wavelength
source free $250 \, \mu$m counterpart map - is then pixel-wise fit to
the long wavelength map to determine the scaling between them.  Bright
sources tend to make such scalings difficult in SPIRE data because
such sources have individual colours which make pixel-wise colour
comparisons have large variance.  However, by construction such bright
sources have been removed from the comparison so the scaling between
the colours is well described by a linear model with coefficients of
$0.76$ at $350 \, \mu$m and $0.40$ at $500 \, \mu$m.  The scaled $250
\, \mu$m bright source removed map is then subtracted from the
equivalent long wavelength map.  This has the effect of subtracting
out both the $250 \, \mu$m sources undetected in the longer wavelength
map and the component of the confused background present in the $250
\, \mu$m map.

As a further step, the $250 \, \mu$m sources whose long wavelength
fits were unsuccessful are binned into a map using their $250 \, \mu$m
flux and position, but the PRF of the longer wavelength channel.  This
undetected source model map is then threshold cut; all pixels with
values greater than the threshold are deemed possible contaminants to
the longer wavelength maps and a pixel mask is generated based on the
positions of the cut sources.  The center and right most panels of
Fig.~\ref{fig:submaps}\ show the resulting $250 \, \mu$m source
removed and masked maps for the $350$ and $500 \, \mu$m channels.

\begin{figure}[h!tb]
\centering
\epsfig{file=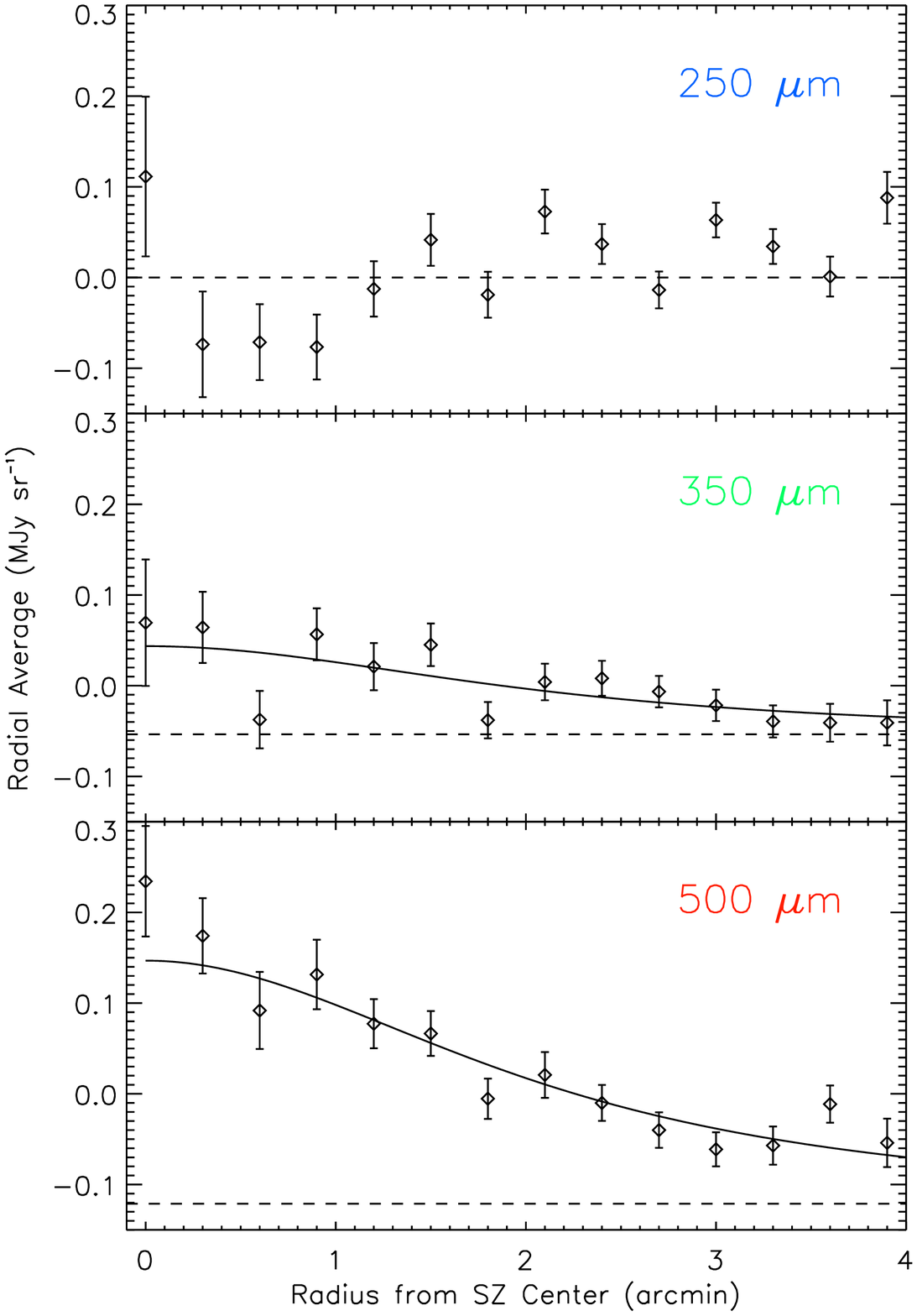,width=0.49\textwidth}
\caption{Radial averages of the source subtracted and masked maps
  shown in Fig.~\ref{fig:submaps}.  As the absolute mean of the maps
  are not measured by SPIRE, the points for each band are scattered
  about $0 \,$MJy sr$^{-1}$.  Also plotted are the best fitting
  isothermal $\beta$ models using the parameters in
  \citet{Halverson2009} in the two SPIRE bands in which the SZ effect
  is expected to be non-zero, $350$ and $500 \, \mu$m; Table
  \ref{tab:results}\ gives the numerical values of the central
  increments brightnesses at both wavelengths.  As the $250 \, \mu$m
  radial bins are not correlated, each point should be compared to the
  (dashed) line $\Delta I = 0 \,$MJy sr$^{-1}$, while in $350$ and
  $500 \, \mu$m the dashed line shows the inferred zero level of the
  SZ effect.}
\label{fig:radave}
\end{figure}

The $500 \, \mu$m map shown in Fig.~\ref{fig:submaps} features an
extended source coincident with the measured center of the SZ effect
in this cluster \citet{Halverson2009}.  If this signal is due to
extended, roughly azimuthally symmetric emission as expected from the
SZ effect then taking radial averages of the masked maps will enhance
the signal to noise ratio of this detection.  As the signal to noise
ratio per map pixel is small, here we assume azimuthal symmetry and
average the maps in simple radial bins of constant 18 arcsec width;
the results of this are shown in Fig.~\ref{fig:radave}.
Interestingly, both the $350$ and $500 \, \mu$m data exhibit extended
emission at the center of the cluster; to constrain this, we fit these
data to the best fitting isothermal $\beta$ model from
\citet{Halverson2009} which has a characteristic angular size of a few
arcminutes.  The resulting fits are plotted in Fig.~\ref{fig:radave};
Table \ref{tab:results}\ lists the best fitting central SZ increment
brightnesses at $350$ and $500 \, \mu$m.  For the
\citet{Halverson2009} model parameters the probability to exceed the
measured $\chi^{2}$ for the fit at $350 \, \mu$m is $0.21$; for $500
\, \mu$m it is $0.25$.  The isothermal $\beta$ model is a poor
statistical description of the radial average at $250 \, \mu$m.  In
addition to the statistical uncertainties associated with the errors
on the data output from the fit, the uncertainties in the
\citet{Halverson2009} model contribute a further $1.2 \,$MJy sr$^{-1}$
at $350 \, \mu$m and $1.9 \,$MJy sr$^{-1}$ at $500 \, \mu$m.  We have
not included the contribution of these in the uncertainties listed in
Table \ref{tab:results}.

\begin{table*}[!htb]
\centering
\caption{Best fitting and expected SZ effect parameters at $350$ and
  $500 \, \mu$m.}
\begin{tabular}{lccc}
\hline
 & \multicolumn{2}{c}{$\Delta I_{0}$ (MJy sr$^{-1}$)} & \\ 
& $500 \, \mu$m & $350 \, \mu$m 
& $\Delta I_{0}(350) / \Delta I_{0}(500)$ \\ \hline\noalign{\smallskip}
SPIRE Best Fit & 
% $26.8 \pm 3.1 \mathrm{(stat.)} \pm 1.9 \mathrm{(model)} \times 10^{-2}$ &  
% $9.7 \pm 1.9 \mathrm{(stat.)} \pm 1.2 \mathrm{(model)} \times 10^{-2}$ \\
 $0.268 \pm 0.031$ & $0.097 \pm 0.019$ & $0.36 \pm 0.12$ \\
Expected with relativistic correction & 
% $26.6 \pm 1.4 \times 10^{-2}$ & $6.0 \pm 0.3 \times 10^{-2}$
 $0.266 \pm 0.014$ & $0.060 \pm 0.003$ & $0.23 \pm 0.05$ \\ 
Expected without relativistic correction & 
% $26.6 \pm 1.4 \times 10^{-2}$ & $6.0 \pm 0.3 \times 10^{-2}$
 $0.184 \pm 0.010$ & $0.017 \pm 0.001$ & $0.09 \pm 0.05$ \\ \hline
\end{tabular}
\label{tab:results}
\end{table*}

\begin{figure}[!htb]
\centering
\epsfig{file=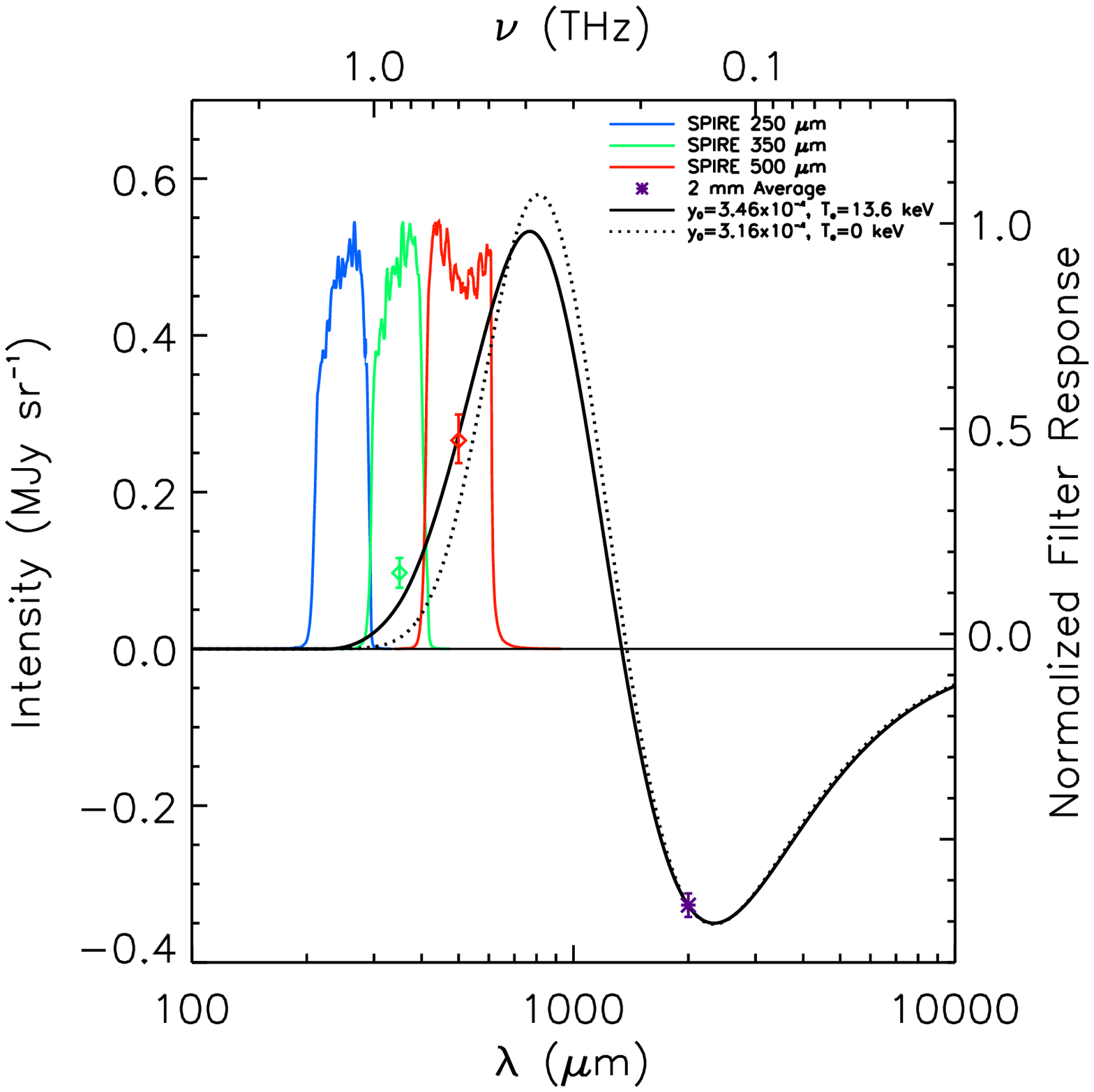,width=0.49\textwidth}
\caption{The SZ effect spectrum in \bulletc.  The $2 \,$mm uncertainty
  weighted average of the measurements of \citet{Andreani1999},
  \citet{Halverson2009} and \citet{Plagge2009} (purple asterisk) leads
  to the SZ spectrum shown (solid black curve).  The best fitting
  SPIRE measurements at $350$ and $500 \, \mu$m (green and red
  diamonds) and the normalized SPIRE bandpasses (blue, green, red
  solid lines) are shown for reference.  The SZ effect spectrum which
  is consistent with the $2 \,$mm measurements but excludes the
  relativistic SZ effect correction is also shown (black dotted line).
  Though both SZ effect curves are well matched in the decrement, the
  finite temperature SZ effect corrections change the results by as
  much as $70$\% of the expected signal in the SPIRE bands.}
\label{fig:SZtemplate}
\end{figure}

In order to compare the SZ effect increments measured here with the
signal expected from previous measurements, we compile the results of
\citet{Andreani1999}, \citet{Halverson2009} and \citet{Plagge2009}.
As all of these measurements were performed with instruments working
at $2 \,$mm, they can be averaged to create a best estimate for the
central Comptonization parameter using the X-ray determined cluster
temperature of $T_{\mathrm{e}}=13.9 \,$keV. Including relativistic
corrections, the uncertainty weighted average $y_{0}$ is $3.46 \pm
0.16 \times 10^{-4}$.  Table \ref{tab:results}\ lists the expected
brightness of the SZ distortion in the SPIRE bands including the
effect of the actual filter bandpasses; Fig.~\ref{fig:SZtemplate}
shows the spectral distortion of the CMB brightness for these $y_{0}$
and $T_{\mathrm{e}}$ values with the SPIRE bandpasses for reference.
Good agreement is found between the expected results given the
fiducial $y_{0}$ and $T_{\mathrm{e}}$ and the SPIRE results at both
$350$ and $500 \, \mu$m.

The relativistic SZ effect corrections required in the presence of a
relativistic electron population have the largest effect at both the
peaks of the SZ effect and the very shortest wavelengths.
\citet{Nozawa2000} has calculated analytic fitting formulae for these
corrections for $\lambda \gtrsim 250 \, \mu$m; in the calculations
above we use these formulae to correct the purely thermal SZ effect
spectrum for the presence of relativistic electrons.  To determine the
change in the SPIRE result if these finite temperature corrections
were absent, we also compute the SZ spectrum without the relativistic
electron correction and central Comptonization parameter which leaves
the brightness at $2 \,$mm constant, $y_{0} = 3.16 \times 10^{-4}$.
Fig.~\ref{fig:SZtemplate}\ shows this pure thermal SZ effect
spectrum and Table \ref{tab:results}\ lists the corresponding $350$
and $500 \, \mu$m SPIRE band weighted intensities.

Table \ref{tab:results}\ also lists the ratio of the $350$ to $500 \,
\mu$m SZ effect intensities for the SPIRE measurement presented here,
the finite temperature SZ effect corrected spectrum, and the pure
thermal SZ spectrum.  The SPIRE and relativistic SZ effect corrected
spectrum are consistent at the $1 \sigma$ level, while the SPIRE
measurement and the purely thermal SZ spectrum are different by $2.1
\sigma$.  This is the strongest evidence for the presence of the
relativistic corrections to date, though not a detection of their
effect on the SZ spectrum.

%________________________________________________________________

\section{Discussion and conclusions}
\label{S:conclusion}

The possibility of detecting the SZ effect significantly shortward of
its positive peak is a testament to the extraordinary capabilities of
SPIRE.  However, works like \citet{Zemcov2007} show that in large
galaxy cluster survey samples, significant contamination to the SZ
signal from bright, gravitationally lensed background sources is
common.  Though the properties of this particular cluster have not
precluded measurement of the SZ effect -- \bulletc\ has fewer bright,
lensed sources close to its SZ effect center than typical clusters
which have been observed in the sub-mm and is relatively bright and
broad in the SZ effect -- based on this single example it is difficult
to determine whether gravitational lensing of the sub-mm background
will make measurements similar to this one more challenging in more
typical clusters at these and other wavelengths.  Data from surveys
like the HLS will allow us to understand whether typical clusters are
suitable for SZ effect increment detection, and how the lensed sub-mm
background will affect measurements at other wavelengths.

The possibility of diffuse emission from cold dust in the ICM has been
discussed in the past (\textit{e.g.} \citealt{Stickel2002}).  It is
expected that, due to sputtering by energetic photons, such dust would
have a very short lifetime in the ICM environment \citep{Draine1979}.
Based on the radial averages of the $250 \, \mu$m source subtracted
maps, where for ICM dust with reasonable temperatures the brightest
thermal emission would occur, we find no evidence for this type of
diffuse emission in this cluster.  Because in any reasonable scenario
such dust emission would be faint in a $z > 0.1$ cluster, we expect
that targeted searches of local clusters will be more successful for
this science, though the HLS and similar surveys can provide useful
constraints.

Given the statistical uncertainties, the $350 \, \mu$m SZ effect
increment measured here is slightly less than $2 \sigma$ larger than
would be expected from the $2 \,$mm measurements.  This may point to
residual problems with the source subtraction procedure.  More data
will allow tuning of the confused sub-mm background removal algorithm
and checks on whether biases arising from poor background removal are
endemic in a large cluster sample.

Given the presence of the SZ effect in clusters, it seems that care
must be taken with photometry of $500 \, \mu$m sources within an
arcminute or so of the cluster center; such sources will be positioned
on a diffuse background so their fluxes will be biased by a small
amount \citep{Rex2010}.

As the peak of the FIR emission of dusty sources is redshifted to
progressively longer wavelengths, high redshift sources are expected
to have exceptionally red spectral energy distributions.  The results
of this work show that care must be taken when searching for sources
based on their presence in the $500 \, \mu$m band alone; galaxy
clusters whose SZ effects are relatively bright and compact could well
masquerade as such sources.  Due to the well known problem of
determining counterparts to sub-mm sources at other wavelengths, such
SZ effect contamination may not be immediately obvious, particularly
as cluster fields are crowded and many possible counterparts may be
present.  As a corollary, searching for very red sources in confusion
limited SPIRE surveys may turn up compact clusters based on the
presence of strong $500 \, \mu$m emission.  The search for such
extremely red SPIRE sources in the HLS and other programmes is
underway now.

%________________________________________________________________

\begin{acknowledgements}

This work is based in part on observations made with Herschel, a
European Space Agency Cornerstone Mission with significant
participation by NASA.  Support for this work was provided by NASA
through an award issued by JPL/Caltech.

\end{acknowledgements}

%________________________________________________________________

\bibliographystyle{aa}
%\bibliography{bullet_diffuse_draft1}
\bibliography{14685}

\end{document}